\documentclass[graybox]{svmult}

\usepackage{mathptmx}       
\usepackage{helvet}         
\usepackage{courier}        
\usepackage{type1cm}        
\usepackage{makeidx}         
\usepackage{graphicx}        
\usepackage{multicol}        
\usepackage[bottom]{footmisc}

\makeindex             

\usepackage{amsmath,amssymb}

\newcommand{\noun}[1]{\textsc{#1}}

\usepackage{url}

\usepackage[margin=2cm]{geometry}

\usepackage{changes}

\begin{document}

\title*{Relating complexities for the reflexive study of complex systems}
\titlerunning{Relating complexities} 
\author{Juste Raimbault}
\institute{Juste Raimbault \at UPS CNRS 3611 ISC-PIF and UMR CNRS 8504 G{\'e}ographie-cit{\'e}s, \email{juste.raimbault@polytechnique.edu}
}
%
%
\maketitle

\abstract{Several approaches and corresponding definitions of complexity have been developed in different fields. Urban systems are the archetype of complex socio-technical systems concerned with these different viewpoints. We suggest in this chapter some links between three types of complexity, namely emergence, computational complexity and informational complexity. We discuss the implication of these links on the necessity of reflexivity to produce a knowledge of the complex, and how this connects to the interdisciplinary of approaches in particular for socio-technical systems. We finally synthesize this positioning as a proposal of an epistemological framework called \emph{applied perspectivism}, and discuss the implications for the study of urban systems.
}



\section{Introduction}

The various disciplines or approaches concerned with the study of urban systems have the common point of having postulated their \emph{complexity}: Batty's ``new science of cities'' is rooted within complexity science paradigms \cite{batty2007cities,batty2013new}; Pumain's evolutive urban theory makes the fundamental assumption of urban systems as co-evolving complex systems \cite{pumain2017geography,pumain1997pour}; Haken's synergetics have application in urban space cognition for example \cite{e18060197}; recent works by physicists have applied tools from statistical physics to urban systems \cite{west2017scale}; architecture has a long tradition in integrating complexity in the urban fabric \cite{alexander1977pattern}; the new economic geography is aware of the complexity of its objects of study despite its reductionism in methods used \cite{krugman1994complex} and has developed an evolutionary branch \cite{cooke2018evolutionary}; artificial life has introduced complex simulation models with application to urban problems \cite{raimbault2014hybrid}; to give a few examples.

What is meant by complexity remains however not well defined, and as \cite{chu2008criteria} recalls, several definitions and characterizations exist and there is a priori no reason to think that these could converge. Interestingly, a reductionist view on this matter would aim at some unified definition, whereas most complexity approaches will take this diversity as an asset to be developed (take for example the complex thinking advocated by \cite{morin1991methode} which relies on the progressive integration of multiple disciplines and thus viewpoints on the world). \cite{manson2001simplifying} develops different notions of complexity and their potential in the study of geographical systems. In the case of urban form and design, \cite{Boeing2018} gives an overview of existing measures of urban form and how they relate to various complexity approaches. The previous chapter \cite{batty2018which} has investigated this issue by providing a broad historical overview of complexity approaches to urban systems, and details several measures of complexity linked for example to fractals and Shannon information. Contrary to \cite{Ladyman2013} which aims at a unified approach of complexity, we will embrace here a diversity of views and highlight their complementarity.

We propose here a theoretical entry to an effective integration of some approaches to complexity, inducing a framework to study complex socio-technical systems. More precisely, we illustrate the links between three types of complexities (emergence, computational complexity, and informational complexity, which choice will be discuss below), and develop the implication of these links on the production of a knowledge of the complex. This work also has more general epistemological implications, since we introduce a development modestly contributing (i.e. in our context of the study of urban systems) to the \emph{knowledge of knowledge} \cite{edgar1986methode}. An implicit aim is thus to interrogate the links between complexity and processes of knowledge production.

Our argument will therefore (i) establish some links between different types of complexities; (ii) explore the necessity of reflexivity possibly as a consequence of these links; (iii) suggest a practical framework to apply these principles to the construction of integrative approaches to complex systems. 

The rest of this chapter is organized as follows: we first introduce more precisely the types of complexities we consider; we then detail some links which are currently not obvious in the literature (i.e. between computational complexity and informational complexity with emergence). After recalling reasons suggesting a necessity of reflexivity, we sketch how the production of knowledge on complex systems falls at the intersection of different complexities and how this implies reflexivity. We finally develop implications for interdisciplinarity, introduce the corresponding applied perspectivism framework and discuss the implications for the study of urban systems.

\section{Complexity and complexities}

What is meant by complexity of a system often leads to misunderstandings since it can be qualified according to different dimensions and visions. We distinguish first the complexity in the sense of weak emergence and autonomy between the different levels of a system, and on which different positions can be developed as in \cite{deffuant2015visions}. We will not enter a finer granularity, the vision of social complexity giving even more nightmares to the Laplace daemon, and since it can be understood as a stronger emergence (in the sense of weak and strong emergence as viewed by \cite{bedau2002downward}).

We thus simplify and assume that the nature of systems plays a secondary role in our reflexion, and therefore consider complexity in the sense of an emergence. This choice answers the rationale that a certain level of complexity can already be present in the \emph{structure} of a system (in the sense of the interaction between its elements) whatever the complexity of its elements themselves. This idea is in line with the ``sociophysics'' approach to socio-technical and human issues as claimed by \cite{caldarelli2018physics}, in the sense of extracting patterns and modeling stylized processes\footnote{We however do not endorse the idea that importing methods and tools from physics to study social problems would consist in a ``physics of human'' since the objects studies are not physical objects anymore - this relates to the debate wether disciplines can be defined by methods or by objects of study; we believe that the emergence of disciplines is far more complex and also involves social phenomena, as illustrated by the imperialist positioning of \cite{caldarelli2018physics}.}. We do not tackle here the question of relating complexities between levels of a system, which remains open (a system composed of complex agents can sometimes be simple and reciprocally) and is out of the reach of this work. This issue is related to finding the appropriate design of levels in modeling and the appropriate ontology to answer a given question \cite{roth2009reconstruction}. \cite{WolfE8678} suggest for example that biological complexity has intrinsic physical roots, implying a transfer of complexity between two distinct levels. 


Beyond the view of complexity as emergence, we distinguish two other ``types'' of complexity, namely computational complexity and informational complexity, that can be seen as measures of complexity, but that are not directly equivalent to emergence. We can for example consider the use of a simulation model, for which interactions between elementary agents translate as a coded message at the upper level: it is then possible by exploiting the degrees of freedom to minimize the quantity of information contained in the message. The different languages require different cognitive efforts and compress the information in a different way, having different levels of measurable complexity~\cite{febres2013complexity}. In a similar way, architectural artefacts are the result of a process of natural and cultural evolution, and witness more or less this trajectory.

Numerous other conceptual or operational characterizations of complexity exist, and it is clear that the scientific community has not converged on a unique definition. Indeed, \cite{chu2008criteria} proposes to continue exploring the different existing approaches, as proxies of complexity in the case of an essentialism, or as concepts in themselves otherwise. This approach that is in a certain way reflexive, since the complexity should emerge naturally from the interaction between these different approaches studying complexity, hence the reflexivity. An example of approach not taken into account here is chaos in dynamical systems, although numerous links for example with computational complexity \cite{PROKOPENKO2019} naturally exist. We propose to focus only on the three concepts described above in particular, for which the relations are already not obvious.

Indeed, links between these three types of complexity are not systematic, and depend on the system considered. Epistemological links can however be introduced. We will develop the links between emergence and the two other complexities, since the link between computational complexity and informational complexity is relatively well explored, and corresponds to issues in the compression of information and signal processing, or moreover in cryptography.

We can furthermore note that complexity is not the only concept for which different approaches diverge for complex enough systems: \cite{thurner2017three} detail three approaches to entropy, linked to information theory, thermodynamics and statistical inference, and show that the corresponding measures do not coincide for some examples of complex systems.

\section{Computational complexity and emergence}

Different clues suggest a certain necessity of computational complexity to have emergence in complex systems, whereas reciprocally a certain number of adaptive complex systems have high computational capabilities.

A first link where computational complexity implies emergence is suggested by an algorithmic study of fundamental problems in quantum physics. Indeed, \cite{2014arXiv1403.7686B} shows that the resolution of the Schrödinger equation with any Hamiltonian is a NP-hard and NP-complete problem, and thus that the acceptation of $\mathbf{P}\neq\mathbf{NP}$ implies a qualitative separation between the microscopic quantum level and the macroscopic level of the observation. Therefore, it is indeed the complexity (here in the sense of their computation) of interactions in a system and its environment that implies the apparent collapse of the wave function, what rejoins the approach of \cite{gell1996quantum} by quantum decoherence, which explains that probabilities can only be associated to decoherent histories (in which correlations have led the system to follow a trajectory at the macroscopic scale). The \emph{Quantum Measurement Problem} arises when considering a microscopic wave function giving the state of a system that can be the superposition of several states, and consists in a theoretical paradox, on the one hand the measures being always deterministic whereas the system has probabilities for states, and on the other hand the issue of the non-existence of superposed macroscopic states (collapse of the wave function). As reviewed by~\cite{schlosshauer2005decoherence}, different epistemological interpretations of quantum physics are linked to different explanations of this paradox, including the ``classical'' Copenhagen one which attributes to the act of observation the role of collapsing the wave function. \cite{gell1996quantum} recalls that this interpretation is not absurd since it is indeed the correlations between the quantum object and the world that product the decoherent history, but that it is far too specific to consider the observer only, and that the collapse happens in the emergence itself: the cat is either dead or living, but not both, before we open the box. The paradox of the Schrödinger cat appears then as a fundamentally reductionist perspective, since it assumes that the superposition of states can propagate through the successive levels and that there would be no emergence, in the sense of the constitution of an autonomous upper level. To summarize, the work of \cite{2014arXiv1403.7686B} suggests that computational complexity is sufficient for the presence of emergence. From a contrapositive point of view, \cite{2017arXiv170404231E} show that quantum computation reduces drastically the memory needed to simulate stochastic processes, highlighting also the role of the memory space dimension in computational complexity. Backing up these ideas, \cite{davies2004emergent} suggests the existence of a minimal computational complexity for a system to exhibit emergence, based on a theoretical information capacity of the universe.

An other approach proposed by \cite{frauchiger2018quantum} demonstrates that quantum theory is inconsistent with its use to describe itself. More precisely, it is shown through a thought experiment that complex macroscopic systems that would answer to quantum rule and would themselves use quantum mechanics to describe the world leads to an intrinsic contradiction. This suggests an other entry to emergence and the decoherence problem, and gives empirical support to this link between emergence and computational complexity.

This effective separation of scales does not a priori imply that the lower level does not play a crucial role, since \cite{vattay2015quantum} proves that the properties of quantum criticality are typical of molecules of the living, without a priori any specificity for life in this complex determination by lower scales: \cite{2016arXiv161102269V} has recently introduced a new approach linking quantum theories and general relativity in which it is shown that gravity could be an emergent phenomenon and that path-dependency in the deformation of the original space introduces a supplementary term at the macroscopic level, which allows explaining the deviations in observational data attributed up to now to \emph{dark matter}.


Reciprocally, the link between computational complexity and emergence is revealed by questions linked to the nature of computation~\cite{moore2011nature}. Cellular automatons, that are moreover crucial for the understanding of several complex systems, have been shown as Turing-complete, i.e. if it is able to compute the same functions than a Turing machine, commonly accepted as all what is ``computable'' (\noun{Church}'s thesis). The Game of Life is such an exemple~\cite{beer2004autopoiesis}. There even exists a programming language allowing to code in the \emph{Game of Life}, available at \url{https://github.com/QuestForTetris}. Its genesis finds its origin in a challenge posted on \emph{codegolf} aiming at the conception of a Tetris game simulated by the game of life, and ended in an extremely advanced collaborative project. This property of the game of life to be used as a computing device can be used to simulate ``meta-pixels'', i.e. a cellular automaton at an upper level \cite{todesco2013cellular}, which behavior weakly emerges from the lower cell but for which rules can be autonomously stated, illustrating a system that could in theory exhibit weak emergence at an arbitrary number of level. This also suggests an importance of reflexivity, on which we will come back below.

Some organisms without a central nervous system are capable of solving difficult decisional problems~\cite{reid2016decision}. An ant-based algorithm is shown by~\cite{Pintea2017} as solving a Generalized Travelling Salesman Problem (GTSP), problem which is NP-difficult. This fundamental link had already been conceived by \noun{Turing}, since beyond his fundamental contributions to contemporary computer science, he studied morphogenesis and tried to produce chemical models to explain it~\cite{turing1952chemical} (that were far from actually explaining it but which conceptual contributions were fundamental, in particular for the concept of reaction-diffusion). We moreover know that a minimum of complexity in terms of constituting interactions in a particular case of agent-based system (models of boolean networks), and thus in terms of possible emergences, implies a lower bound on computational complexity, which becomes significant as soon as interactions with the environment are added~\cite{tovsic2017boolean}.



\section{Informational complexity and emergence}


Informational complexity (see \cite{dedeo2016information} for a smooth introduction to information theory), or the quantity of information contained in a system and the way it is stored, also bears some fundamental links with emergence. Information is equivalent to the entropy of a system and thus to its degree of organisation - this what allows to solve the apparent paradox of the Maxwell Daemon that would be able to diminish the entropy of an isolated system and thus contradict the second law of thermodynamics: it indeed uses the information on positions and velocities of molecules of the system, and its action balances to loss of entropy through its captation of information (the Maxwell Daemon is more than an intellectual construction: \cite{cottet2017observing} implements experimentally a daemon at the quantic level).

This notion of local increase in entropy has been largely studied by \noun{Chua} under the form of the \emph{Local Activity Principle}, which is introduced as a third principle of thermodynamics, allowing to explain with mathematical arguments the self-organization for a certain class of complex systems that typically involve reaction-diffusion equations~\cite{mainzer2013local}.

The way information is stored and compressed is essential for life, since the ADN is indeed an information storage system, which role at different levels is far from being fully understood. Cultural complexity also witnesses of an information storage at different levels, for example within individuals but also within artefacts and institutions, and information flows that necessarily deal with the two other types of complexities. Information flows are essential for self-organization in a multi-agent system. Collective behaviors of fishes or birds are typical examples used to illustrate emergence and belong to the canonic examples of complex systems. We only begin to understand how these flows structure the system, and what are the spatial patterns of information transfer within a \emph{flock} for example: \cite{crosato2017informative} introduce first empirical results with transfer entropy for fishes and lay the methodological basis of this kind of studies. Similarly, \cite{lecheval2018social} show that nonlinear interaction between fishes are essential for the propagation of information in the school during a collective U-turn. \cite{lizier2008local} introduces a local measure of transfer entropy to characterize information diffusion in cellular automatons. In the field of artificial life, open-ended evolution is an open research area closely link to emergence, and \cite{corominas2018zipf} introduce a definition of open-ended evolution based on algorithmic information theory.

Furthermore, different theoretical approaches of complex systems suggest a strong link between informational complexity and self-organization. For example, \cite{e18060197} develops an overview from the point of view of synergetics on the role of different type of informations in self-organizing systems ranging from neuroscience and cognition to urban dynamics. The theoretical approach of multi-scale information to complex systems proposed by \cite{allen2017multiscale}, defining an information profile across scales which shape will be linked to the complexity of the system, is an other crucial entry into complexity from the viewpoint of information theory. \cite{gershenson2012world} proposes to interpret complex systems as evolving information, and introduces rules for the behavior of information which are particularly suited to the understanding of cognition and life. Finally, \cite{hoel2017map} shows that the causal structure of some systems in terms of information theory can only be captured by considering emerging levels.

\section{Reflexivity in the study of complex systems}

Furthermore, one aspect of knowledge production on complex systems which seems to be recurrent and even inevitable, is a certain level of reflexivity (and that would be inherent to complex system in comparison to simple systems, as we will develop further). We mean by this term both a practical reflexivity, i.e. a necessity to increase the level of abstraction, such as the need to reconstruct in an endogenous way the disciplines in which a reflexion is positioned as proposed by \cite{2017arXiv171200805R}, or to reflect on the epistemological nature of modeling when constructing a model, but also a theoretical reflexivity in the sense that theoretical apparels or produced concepts can recursively apply to themselves.

The practical inevitability of reflexivity is well-known in social sciences and humanities, but \cite{bourdieu2004science} postulated this would be more generally linked to the nature of scientific knowledge which is inherently social \cite{maton2003reflexivity}. In the case of a strong constructivist approach to science, knowledge is by nature reflexive but highly contingent to the social structure producing it. The question remains open for disciplines in which the object of study is not social or anthropologic, but as soon as producers of knowledge are potential subjects to be studied, the discipline becomes reflexive. There are some clues that reflexivity may be more generally needed in the study of complex systems and not only social systems: in the context of simulating open-ended evolution, what is a grand challenge in the field of artificial life, \cite{banzhaf2016defining} suggest that a reflexive programming language, in the sense that it can write or embed parts of its own code recursively, would be a crucial feature to simulate open-ended evolution.

This practical observation can be related to old epistemological debates questioning the possibility of an objective knowledge of the universe that would be independent of our cognitive structure, somehow opposed to the necessity of an ``evolutive rationality'' implying that our cognitive system, product of the evolution, mirrors the complex processes that led to its emergence, and that any knowledge structure will be consequently reflexive. We naturally do not pretend here to bring a response to such a broad and vague question as such, but we propose a potential link between this reflexivity and the nature of complexity.

\section{Production of knowledge}

\subsection{From complexity to reflexivity}

We know have enough material to come to the importance of reflexivity in studying complexity. It is possible to position knowledge production at the intersection of interactions between types of complexity developed above. First of all, knowledge as we consider it can not be dissociated from a collective construction, and implies thus an encoding and a transmission of information: this relates at an other level to all issues linked to scientific communication. The production of knowledge thus necessitates this first interaction between computational complexity and informational complexity. The link between informational complexity and emergence is introduced if we consider the establishment of knowledge as a morphogenetic process. It is shown by \cite{antelope2016interdisciplinary} that the link between form and function is fundamental in psychology: we can interpret it as a link between information and meaning, since semantics of a cognitive object can not be considered without a function. \cite{hofstadter1980godel} recalls the importance of symbols at different levels for the emergence of a thought, that consist in signals at an intermediate level. Finally, the last relation between computational complexity and emergence is the one allowing us a positioning in particular on knowledge production on complex systems, the previous links being applicable to any type of knowledge.

Therefore, any \emph{knowledge of the complex} embraces not only all complexities and their relations in its content, but also in its nature as we just showed. The structure of knowledge in terms of complexity is analog to the structure of systems its studies. We postulate that this structural correspondence implies a certain recursivity, and thus a certain level of \emph{reflexivity} (in the sens of knowledge of itself and its own conditions). In other words and as we will detail more below, understanding the complex requires complexity, what is intrinsically reflexive.

%

\subsection{The complexity of interdisciplinarity}

We can try to extend to reflexivity in terms of a reflexion on the disciplinary positioning: following \cite{pumain2005cumulativite}, the complexity of an approach is also linked to the diversity of viewpoints that are necessary to construct it. Some links with the previous types of complexities naturally appear: for example, \cite{gell1995quark} considers the effective complexity as an \emph{Algorithmic Information Content} (close to Kolmogorov complexity) of a Complex Adaptive System \emph{which is observing an other} Complex Adaptive System, what gives their importance to informational and computational complexities and suggests the importance of the observational viewpoint, and by extension of their combination. This furthermore paves the way for the perspectivist approach of complex sciences we will introduce below. 

This ``interdisciplinary complexity'' would be a supplementary dimension linked to the knowledge of complex systems. To reach this new type of complexity, reflexivity must be at the core of the approach. \cite{read2009innovation} recall that innovation has been made possible when societies reached the ability to produce and diffuse innovation on their own structure, i.e when they were able to reach a certain level of reflexivity. The \emph{knowledge of the complex} would thus be the product and the support of its own evolution thanks to reflexivity which played a fundamental role in the evolution of the cognitive system: we could thus suggest to gather these considerations, as proposed by \noun{Pumain}, as a new epistemological notion of \emph{evolutive rationality}.

It is highly likely that these approaches could not be tackled in a simple way. Indeed, we can remark that given the law of \emph{requisite complexity}, proposed by \cite{gershenson2015requisite} as an extension of \emph{requisite variety}~\cite{ashby1991requisite}. One of the crucial principles of cybernetics, the \emph{requisite variety}, postulates that to control a system having a certain number of states, the controller must have at least as much states. \cite{gershenson2015requisite} proposes a conceptual extension to complexity, which can be theoretically justified for example by \cite{allen2017multiscale} which introduce the multi-scale \emph{requisite variety}, showing its compatibility with a theory of complexity based on information theory. Therefore the \emph{knowledge of the complex} will necessarily have to be a \emph{complex knowledge}. This other point of view reinforces the necessity of reflexivity, since following \noun{Morin} (see for example \cite{morin1991methode} on the production of knowledge), the \emph{knowledge of knowledge} is central in the construction of a complex thinking.

\section{Discussion}

\subsection{Towards an epistemological framework of applied perspectivism}

The view of complexity we just gave can be formulated as a research program, for the development of an \emph{applied perspectivism}, which has already been sketched for example by \cite{2018arXiv180807282B}. This epistemological positioning is based on Giere's perspectivism \cite{giere2010scientific} and aims at sketching practical research guidelines from its viewpoint. We recall that perspectivism has been commented as a ``third way'' beyond the constructivism-realism debate, which stipulates that any scientific knowledge construction process as a perspective by an agent to answer a purpose with a media, which is called a model. This cognitive approach to science \cite{giere2010explaining} is compatible with the view of cognitive systems by \cite{gershenson2012world} as agents processing information and with the view of \cite{gell1995quark} of effective complexity as already mentioned.

This approach is particularly relevant for the study of complex multidimensional systems: \cite{muelder2018} show for example that even different formalizations of the same theory can lead to highly different outcomes. This does not imply that one is more relevant than others (this point being validated by the internal consistence and external validations), but on the contrary that they inform on different dimensions of the system.

\cite{raimbault2017applied} has introduced a knowledge framework to study complex systems, based on complementary \emph{knowledge domains}, which are the theoretical, empirical, modeling, data, methods and tools domains. Any scientific perspective on a complex system would then be a co-evolutive system of cognitive agents and the knowledge domains. Applied perspectivism would then consist in the reflexive quantification of these dynamics during the production of knowledge itself, and a proactive engagement of agents in the explicit positioning of their perspectives and the construction of integrative coupled perspectives.

The main principles of applied perspectivism would in particular include, echoing several of the golden rules by \cite{banos2013pour} for simulation models in social sciences:

\begin{itemize}
	\item Actively foster the consistence of several perspectives when studying a common object, and foster their communication and integration. This point is exactly the virtuous spiral between disciplinarity and interdisciplinarity introduced by \cite{banos2013pour}.
	\item To ease the coupling of perspectives, each view to be included must be well self-aware of its own positioning in the scientific landscapes, of its own strengths and weaknesses, and what it can bring to an integrated perspective. To achieve this, an increased reflexivity of each discipline, including the ones which are traditionally not reflexive by nature, is crucial.
	\item The new simulation model exploration methods, which have recently provided new ways of producing knowledge \cite{pumain2017urban}, help for a higher integration of knowledge domains, and thus of the components of perspectives. This is in line of considering computer simulation as genuine experiments themselves \cite{boge2019computer}.
	\item Assuming some kind of ``transfer hypothesis'' between the models and the perspectives of which they are the media, an approach to couple perspectives (and for example theories) is achieved through the coupling of corresponding models. How to deal with possible ontological incompatibilities and how to couple models in general are open research questions that still need to be investigated for this point to be relevant.
\end{itemize}

At this stage, this framework remains at a proposal stage, and several directions must be developed to make it more robust: (i) as the framework acts in a way as a ``model of scientific activities'', we believe that a formalization into a modeling or logic framework would ensure a greater consistence (although several approach would be possible, recalling the reflexivity through the recursion in the framework itself that would need several perspectives to be described); (ii) in that spirit, specific simulation models can be developed to answer specific questions such as finding the appropriate compromise in the disciplinary-interdisciplinary interplay, or including more social aspects in the framework (incentives, relations between individuals); (iii) following these preliminary experiments, different practical implementations could be proposed and tested on real cases, such that these experimentations would inform back the framework itself and the different models.

In the concluding chapter of this book \cite{pumain2019perspectives}, a sketch of reflexive analysis of urban theories is proposed, through a citation network analysis of the scientific neighborhood of cited references of all chapters in the book. The resulting knowledge maps provide an overview of involved disciplines and their relative positioning. Each approach could then use this knowledge on itself to refine research directions or interactions with other approaches.

\subsection{Implications for urban theories}

Our tour on complexities and reflexivity has direct implications for urban theories, since as we already developed, this kind of socio-technical system is the archetype of a complex system on which numerous complementary views can be formulated. We suggest that these aspects may transfer to the models themselves if they become relatively independent: \cite{white2017necessity} speculates that future urban modeling approaches will have to be themselves self-adaptive and be based on artificial intelligence to dynamically develop their responses to changing contexts of application. The development of multi-scale territorial models for sustainable policies, recalled as a crucial issue by \cite{rozenblat2018conclusion}, will indeed necessarily require first a practical epistemological reflexivity to integrate the several disciplines concerned (in particular geography and economics which have still much to do to communicate \cite{raimbault2017invisible}), and secondly a reflexion on the emergence of intelligence in territorial systems themselves.

Our work suggests the crucial role of reflexivity to study urban systems, and some links remain also to be established with other types of reflexivity and methods to reach it, such as the one proposed by \cite{anzoise2017perception} to enhance reflexivity of all actors in anthropological research, including not only researchers but also others stakeholders and the people studied. Also, our view of integrated knowledge domains, goes as already suggested by \cite{raimbault2017cadre} in a direction beyond the arbitrary ``qualitative-quantitative'' opposition which is a recurrent issue in social sciences, and echoes previous proposals in the literature such as \cite{shah2006building} suggesting that more robust theories shall get rid of this opposition.

Finally, we can try to summarize from these epistemological developments some  ``practical'' implications that could be used as useful guides for complex urban research:

\begin{enumerate}
	\item As suggested by the applied perspectivist positioning, the coupling of models should play a crucial role in the capture of complexity.
	\item Similarly, the inspiration for complex urban approaches shall essentially be interdisciplinary and aim at combining different points of view.
	\item Different knowledge domains can not be dissociated for any complex approach, and should be use them in a strongly dependent way.
	\item These approaches should imply a certain level of reflexivity.
	\item The construction of a complex knowledge \cite{morin1991methode} is neither inductive nor deductive, but constructive in the idea of a morphogenesis of knowledge: it can be for example difficult to clearly identify precise ``scientific deadlocks'' since this metaphor assumes that an already constructed problem has to be unlocked, and even to constrain notions, concepts, objects or models in strict analytical frameworks, by categorizing them following a fixed classification, whereas the issue is to understand if the construction of categories is relevant. Doing it a posteriori is similar to a negation of the circularity and recursion of knowledge production. The elaboration of ways to report knowledge that would translate the diachronic character and its evolutive properties remains an open problem.
\end{enumerate}

\section*{Conclusion}

We started a journey into concepts of complexity by illustrating numerous approaches from diverse fields considering urban systems as complex and introducing corresponding concepts. This suggested to investigate the links between different types of complexities. We then suggested that the production of knowledge on the complex is at the intersection of these links between complexities, and hence that reflexivity of this knowledge could be necessary. This naturally brought us to complexity as a diversity of viewpoints on a system. These considerations are summarized into a proposal of an epistemological framework of \emph{applied perspectivism}, that we believe to be a useful tool for future complex urban theories and models. A broader integration with other views of complexity and existing approaches in social sciences remains to be done, such as evidence-based foundations of the approach, achieving its own reflexivity.




\end{document}